*Review Article*

# Survey of Promising Technologies for 5G Networks

**Nam Tuan Le,[1] Mohammad Arif Hossain,[1] Amirul Islam,[1] Do-yun Kim,[2] Young-June Choi,[2] and Yeong Min Jang[1]**

[1]*Department of Electronics Engineering, Kookmin University, Seoul, Republic of Korea*
[2]*Department of Computer Engineering, Ajou University, Suwon, Republic of Korea*

Correspondence should be addressed to Yeong Min Jang; yjang@kookmin.ac.kr

As an enhancement of cellular networks, the future-generation 5G network can be considered an ultra-high-speed technology. The proposed 5G network might include all types of advanced dominant technologies to provide remarkable services. Consequently, new architectures and service management schemes for different applications of the emerging technologies need to be recommended to solve issues related to data traffic capacity, high data rate, and reliability for ensuring QoS. Cloud computing, Internet of things (IoT), and software-defined networking (SDN) have become some of the core technologies for the 5G network. Cloud-based services provide flexible and efficient solutions for information and communications technology by reducing the cost of investing in and managing information technology infrastructure. In terms of functionality, SDN is a promising architecture that decouples control planes and data planes to support programmability, adaptability, and flexibility in ever-changing network architectures. However, IoT combines cloud computing and SDN to achieve greater productivity for evolving technologies in 5G by facilitating interaction between the physical and human world. The major objective of this study provides a lawless vision on comprehensive works related to enabling technologies for the next generation of mobile systems and networks, mainly focusing on 5G mobile communications.

## 1. Introduction

Mobile communication and wireless networks have advanced phenomenally during the last decade. The ever-growing increase in the demand for resources, especially for multimedia data, with high quality of service (QoS) requirements, has promoted the development of 3G and 4G wireless networks. Nevertheless, the achievements of the development in technology cannot fulfill the proper satisfaction. Therefore, the idea of 5G networks that represent networks beyond 4G has become the need of the hour. 5G networks have come into existence owing to the numerous challenges facing 4G networks, such as need for higher data rate and capacity, lower cost, lower end-to-end latency, and massive inter-device connectivity. However, a comprehensive analysis of future networks or next generation networks of information systems that discusses in related forums and standardization is really challenging. The enabling technologies for next generation mobile systems and networking have been surveyed in this paper, which provides readers a clear vision of the current status.

The planning for future network architecture seems to be the definition of Next Generation Networks (NGN). NGN, a great issue for the internet protocol- (IP-) based future of mobile network infrastructure, is considered as a convergence of communication networks which tries to reduce cost and offers integrated services via a core backbone network. It inherits three different advantages of various networking technologies, namely, layered structure, standard interfaces and multiple services, and functions that can be implemented in several layers ranging from MAC to application. With the increase in the number of Internet users and QoS requirements, NGN has become a moving trend for deployment. It established convergence of user access and integrated communication network services with IP technology. The motivation behind the migration of networking systems from the traditional telecommunication network to NGN has been developed based on the advantages of backbone cost reduction, possibility of fast and new service deployment, controllable QoS, compatibility between fixed and wireless networks, network management centralization, and so on. Existing network services based multimedia application such as voice, data, and video transmission at high speeds will be offered as an important outcome of NGN deployment for the fixed and mobile service integration topology. Furthermore, NGN provides low-cost service at high data rates.

The concept of the future network can also be the fifth-generation mobile system, 5G. Over the course of the long history of mobile communication systems from the first generation to 4G LTE-A (Long Term Evolution Advanced), the mobile communications industry has achieved enormous advances in data communication. The next generation can be a revolution in mobile networks that will achieve the best performance in terms of coverage capability, energy consumption, data speeds of 1 Gbps, and better security and energy efficiency over spectral compared to previous networking systems. However, the next generation wireless communication network has not been defined and characterized exactly. Research on 5G has been initiated by many projects, organizations, and standardization forums. Such research on 5G might be directed by the limitations of current technologies. The key requirements of 5G are real wireless communication with no limitation of coverage edge, access policy, and density zone. Secondly, the network should be able to support high-resolution multimedia (HD) broadcasting service. Thirdly, it should have faster data speeds than the previous generations. Finally, it should support new services based on wearable devices. In addition, the NGN is expected to have massive interdevice connections, which can be termed as Connection of Things. The research on 5G is different from that on previous-generation networks because of the limitations of resources in the RF band. The 5G wireless network will mainly focus on new spectrum, multiple-input-multiple-output (MIMO) diversity, transmission access, and new architecture for capacity and connection time [1].

It is a very challenging issue to meet the QoS requirement at a selection service for any network architecture. The convergence of networking and cloud computing are under the consideration in NGNs to cope with the QoS demand. The controllability, management, and optimization of computing resources are the main factors affecting networking performance in the case of cloud computing. One of the advantages of cloud computing is that it is encapsulation-free, which means that users can access services from any location irrespective of host or end device. User can use services without understanding how they operate or deliver data. However, large numbers of vendors are getting interested in information support, storage, and resource computation using cloud-hosting services. With traditional web technology-based services, the relative positions of client and server strongly affect the system QoS and the quality of experience (QoE) [2]. Therefore, future-generation wireless networks are faced with multiple emerging challenges. The Internet of things (IoT) has emerged as one of the leading technologies for future-generation technologies because it is based on the concept of device interconnection, which can be a step toward achieving the QoS and QoE requirements. It is a conceptualization of a cyber-physical system (CPS), a way for using embedded technologies in the future-generation network. Physical systems are unified with the networking and computation system. The scalability of the future-generation network depends on the IoT system because it is a method of assisting connections among a large number of devices in a whole system. IoT has evolved as a system of uninterrupted communication between any device with another device at any place and time. However, the IoT architecture has come under question lately because it is very difficult to support all devices in an inflexible architecture with the traditional networking system. Consequently, several organizations, companies, and committees are working on the standardization issues of IoT to create a unique platform for future-generation networks.

The development of networking depends on the flexibility and mobility of users, and server visualization, which plays an important role in responding effectively and in a timely manner to the dynamic requirements of applications or users. The traditional network infrastructure is continuously becoming obsolete because of the lack of these features. Moreover, manual changes in network configurations increase the complexity of network management, making it nearly impossible at times. The existing infrastructure cannot support priority-based packet-forwarding or dynamic resource allocation to users. Hence, network management, at its root level, has become a challenging issue owing to the limitations of traditional hardware-based networking, such as complex and costly network configuration, and lack of policy changes and fault management. As networking technologies evolve, the network should be able to support the ever-changing networking functionalities of future network infrastructure, such as integration with new services, dynamic network control, better QoS, and efficient packet-forwarding. However, traditional networks cannot support the ever-changing demand of networking technologies. Therefore, software-defined networking (SDN), an emerging technology, can be employed to overcome the limitations of the current networks with the separation of network control from the underlying data planes or switching devices. By breaking the virtual integration between the data plane and the control plane and by using a centralized SDN controller, SDN provides flexibility in changing network policies, easy hardware implementation, and facilitates network innovation and evolution [3, 4]. By integrating SDN with network function virtualization (NFV), one can gain a global view of the entire network by using an open interface such as OpenFlow and the centralized network controller. SDN can support new services and programs at any level of user requirement or need. Furthermore, SDN has attracted considerable interest from both academia and industry over the past few years. It is, after all, an important step in the evolution and development of future network infrastructures.

In this paper, we provide a comprehensive overview of the ongoing research on the enabling technologies for the 5G network. We present the status of work on the important technologies and service models for the next generation of mobile systems and networks. The remainder of this paper is organized as follows. A new model for network control, SDN, and NFV is described in Section 2, while Section 3 presents a survey of the cloud computing model from the viewpoints of network operation and management. The current standardization status, architectures, and applications of IoT for 5G networks are discussed in Section 4. An overview of mobile access networks is presented in Section 5. Our concluding remarks are given in Section 6.

## 2. Software-Defined Networking (SDN) for 5G

### 2.1. SDN and NFV

*2.1.1. Software-Defined Networking (SDN).* Software-defined Networking (SDN) has been introduced for data networks and next generation Internet [5–8]. It has been defined in several ways. The most unambiguous and established definition is provided by the Open Networking Foundation (ONF) [9, 10], a public association dealing with the standardization, development, and commercialization of SDN. The definition is as follows:

> *"Software-Defined Networking (SDN) is an emerging architecture that is dynamic, manageable, cost effective, and adaptable, where control is decoupled from data forwarding and the underlying infrastructure, and directly programmable for network services and applications".*

According to this definition, SDN has the following characteristics: (i) it decouples network control from the underlying data plane (i.e., switches and routers); (ii) it allows the control plane to be programmed directly through an open interface, for instance, OpenFlow [11, 12]; and (iii) it uses a network controller, (i.e., SDN controller) to define the behavior and operation of the networking infrastructure. SDN can be an ideal prospective for the high-bandwidth, dynamic nature of network management. SDN provides the flexibility to change the network configuration at the software level, thus reducing the necessity of modification at the hardware level. SDN makes it easier to introduce and deploy new applications and services than the traditional hardware-operated networking architectures. It also ensures the QoS at any level of user requirement. Consequently, it will be an attracting architecture from the viewpoint of reconfiguring and redirecting complex networks for real-time management.

*2.1.2. Network Function Virtualization (NFV).* An important observation of SDN is NFV [13]. SDN and NFV are mutually beneficial, but they are not fully dependent on each other. In fact, network functions can be employed and virtualized without using an SDN and vice versa. As it is complementary to SDN, NFV can effectively decouple network functionalities and implement them in software. Thus, it can decouple network functions, for instance, routing decisions, from the underlying hardware devices such as routers and switches, and centralize them at remote network servers or in the cloud through an open interface such as OpenFlow. Hence, the overall network architecture can be highly flexible for fast and adaptive reconfiguration.

The combined functionalities of SDN and NFV [14] make SDNs more advantageous than traditional hardware-based networks. The main advantages can be listed as follows: cost minimization, reduced power consumption through equipment consolidation, reduced processing time by minimizing the typical network operator cycle of innovation, centralized network provisioning by decoupling the data plane from network control plane, extension of capabilities, hardware savings, cloud abstraction, guaranteed content delivery, physical versus virtual networking management, and so on. The advantages of SDN are well explained in [15]. Figure 1 shows a comparison between conventional hardware-based networks and SDN. Furthermore, the differences between SDN and conventional hardware-based networks are summarized in Table 1.

Table 1: Differences between SDN and conventional networking.

| Software-defined networking | Conventional hardware-based networking |
| --- | --- |
| Data and control plane are decoupled by API or OpenFlow | Data and control plane are mounted on same plane, new protocol for every service |
| Automatic reconfigurable and replicing logically centralized configuration | Static or manual configuration and reconfiguration takes time |
| SDN can prioritize or block specific packets | Conventional network leads all packets the same way |
| Provides global or comprehensive network views leading to consistent and effective policies | Provides limited information about networks |
| Easy to program according to application and user needs and can be developed quick via software upgrades | Difficult to replace the existing program with new ideas and works according to packet-forwarding tables |

### 2.2. SDN Functionalities.
SDN can support multiple functionalities because of its centralized controller and separated data and control plane. The SDN's functionalities, along with its layers and planes, are shown in Figure 2. The general functionalities of SDN are as follows.

*Programmability.* Network control is directly programmable as the control plane is decoupled from the forwarding or data plane. SDN allows the control plane to be programmed using different software development tools along with the function of customization of the control network according to user requirements.

*Centrally Managed.* In an SDN, the controller network is logically centralized, thus providing a comprehensive view of the network that appears to the applications or users as a logical device.

*Flexibility.* SDN provides flexibility to network managers. Network managers can manage, configure, secure, and optimize network parameters very rapidly through dynamic, automated SDN programs. This helps the controllers respond to traffic variations. As controllers run in software, SDN affords the flexibility of synchronization through the network

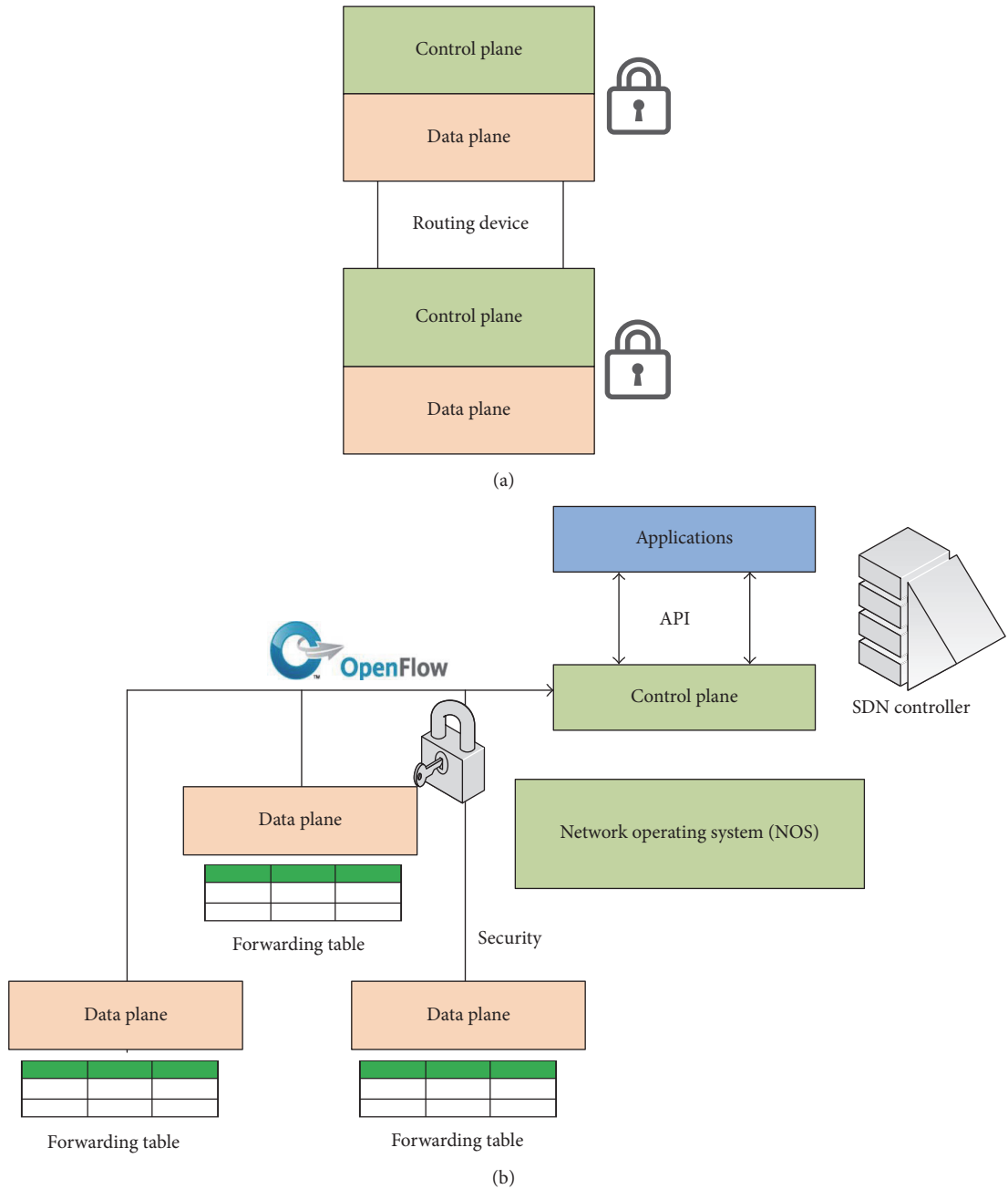

FIGURE 1: Comparison between (a) traditional hardware-based network and (b) SDN.

operating system (NOS) approach on different physical or virtual hosts.

*Granularity*. Since networking is spreading across different protocol layers and the level of data flow is aggregating as well, SDN has the features to control the traffic flow with different granularity on the protocol layers and at the aggregate level. These can vary from the core networks to a single connection in a home LAN.

*Protocol Independence*. SDN has a key feature called protocol independence. It helps run or control a variety of networking protocols and technologies on different SDN network layers. It also enables one to change policies from old to new technologies and supports different protocols for different applications.

*Open Standard-Based*. Instead of multiple vendor devices and protocols, SDN controllers simplify network operation and design based on controller instructions applied through an open standard.

*Ability of Dynamic Control*. SDN has the ability to modify the network traffic flow dynamically. Dynamic reconfiguration covers wide-area networks, and in data center networks, where constant or continuous transportation of real or virtual machines and their network control schemes need to change in minutes or even seconds.

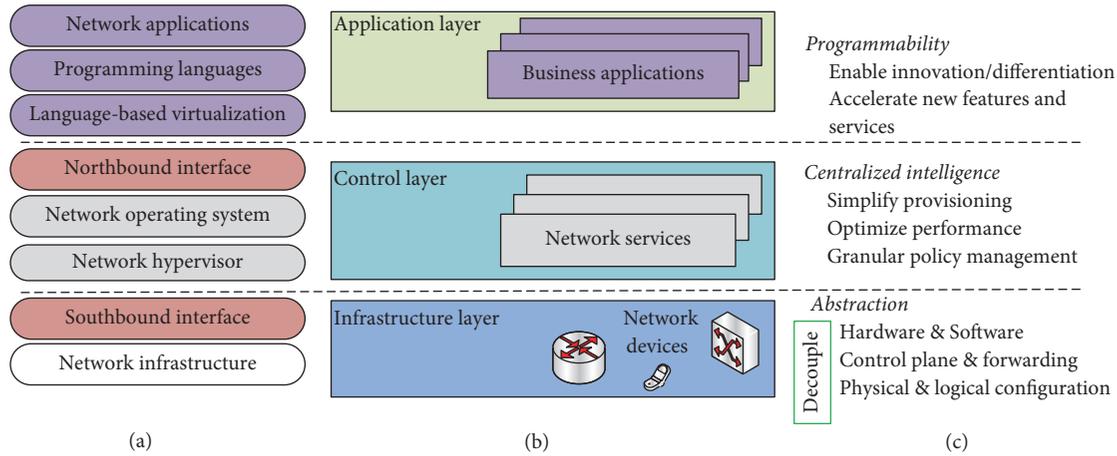

FIGURE 2: SDN (a) layers, (b) planes, and (c) functionalities.

### 2.3. SDN Architecture for 5G.
ONF proposed a simple high-level architecture for SDN. This model can be separated into three layers, namely, an infrastructure layer, a control layer, and an application layer, assembled over each other, as shown in Figure 2(b) [9]. These three layers are described below.

The *infrastructure layer* mainly consists of forwarding elements (e.g., physical and virtual switches, routers, wireless access points) that comprise the data plane. These devices are mainly responsible for (i) collecting network status, storing them temporally in local network devices and sending the stored data to the network controllers and (ii) for managing packets based on the rules provided by the network controllers or administrators. They allow the SDN architecture to perform packet switching and forwarding via an open interface.

The *control layer*, also known as control plane, maintains the link between the application layer and the infrastructure layer through open interfaces. Three communication interfaces allow the controller to interact with other layers, namely, the southbound interface for interacting with the infrastructure layer, northbound interface for interacting with the application layer, and east/westbound interfaces for communicating with groups of controllers. Their functions may include reporting network status and importing packet-forwarding rules and providing various service access points in various forms.

The *application layer* is designed mainly to fulfill user requirements. It consists of the end-user business applications that consume network services. SDN applications are able to control and access switching devices at the data layer through the control plane interfaces. SDN applications include network visualization, dynamic access control, security, mobility and migration, cloud computing, and load balancing (LB). Figure 3 shows the overall architecture of SDN for the 5G mobile system. The details of SDN layers are explained below.

### 2.3.1. Infrastructure Layer.
The underlying infrastructure layer in SDN consists of switching devices that are interconnected to communicate in a single physical network. In SDN, these forwarding devices are generally represented as basic forwarding hardware or device. These devices are connected wirelessly, using optical fibers, optical wires, cloud networks, and so forth. They maintain connection with the controller through an open interface known as the southbound interface. In most SDNs, OpenFlow is used as the open southbound interface. OpenFlow is a flow-oriented protocol and has switches and port abstraction for flow control.

*OpenFlow.* The OpenFlow protocol maintained by ONF [19] is a fundamental element for developing SDN solutions and can be treated as an encouraging consideration of any networking abstraction. OpenFlow, the first leading authorized communications interface linking the forwarding and controls layers of the SDN architecture, allows manipulation and control of the forwarding plane of network devices (e.g., switches and routers) both physically and virtually. OpenFlow helps SDN architecture to adapt to the high-bandwidth, dynamic nature of user applications, adjust the network to different business needs, and interestingly reduce management and maintenance complexity. Figure 4(a) shows the model of the OpenFlow protocol whereas the algorithm is shown in Figure 4(b). When a new flow or packet reaches, some lookup manner originates in the primary lookup table and concludes either with a match in the flow tables or with an error depending on the rules specified by the controller. When the packets do not acknowledge what to do with a distinct incoming packet, default information to forward the packet to the controller is "send to controller" in the case of any unmatched entry. If a link or port change is triggered, event-based messages are sent by forwarding devices to the controller.

Once the rules are matched with the flow rules, the rule's counter is incremented and actions based on the set rules start getting executed. This could lead to forwarding of a packet, after modifying some of its header fields to a specific port or (i) dropping of the packet and (ii) reporting of the packet back to the controller. The summary some of the most significant characteristics of the data plane [20]. However, OpenFlow is not the only available southbound interface for SDNs. There are other API proposals such as Forwarding and Control Element Separation (ForCES) [21]; Open vSwitch Database (OVSDB) [22]; Protocol-oblivious Forwarding (POF) [23, 24]; OpFlex [25]; OpenState [26]; Revised OpenFlow Library (ROFL) [27]; Hardware Abstraction Layer (HAL) [28, 29]; and Programmable Abstraction of Data path (PAD) [30].

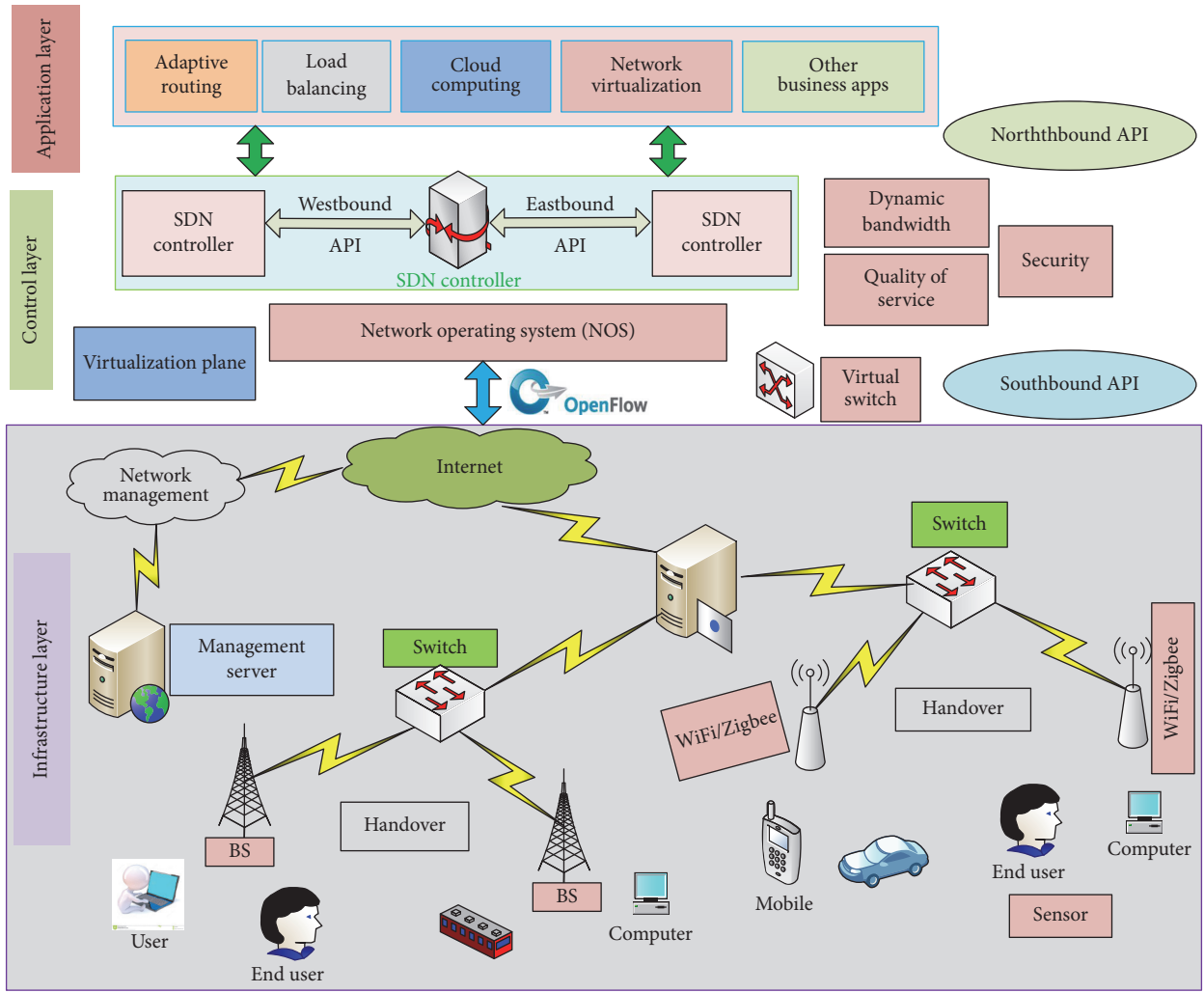

FIGURE 3: SDN architecture for 5G.

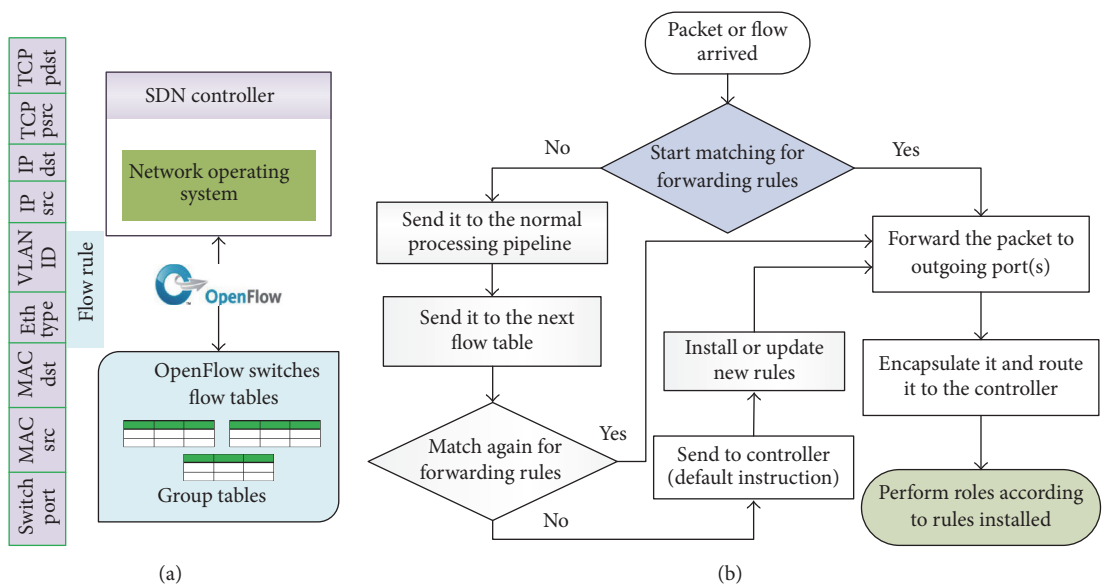

(a)

(b)

FIGURE 4: (a) OpenFlow model and (b) detailed process of OpenFlow protocol.

*2.3.2. Network Controller or Network Operating System (NOS).* The network controller, SDN controller or NOS, is the heart of SDN architecture. It lies between network devices and applications. It is based on operating systems in computing. In [31], the controller is described as software abstraction that controls all functionalities of any networking system. It maintains control over the network through three interfaces, namely, southbound interface (e.g., OpenFlow), northbound interface (e.g., API), and east/westbound interfaces. The southbound interface abstracts the functionalities of programmable switches and connects them to the controller. The northbound interface [32] allows high-level policies or network applications to be deployed easily and transmits them to the NOS, while the east/westbound interfaces maintain communications between groups of SDN controllers. Thus far, many SDN controllers have been proposed by researchers to facilitate controller functionalities. For example, NOX [33] is the first, publicly available OpenFlow controller implementation that can run in Windows, Linux, Mac OS, and other platforms; an extension of NOX has been implemented in POX [34], which is a real Python-based controller; a Java-based controller implementation is called Beacon [35], while Floodlight controller is an extension of Beacon [36], and so on.

The *functionalities* of an SDN controller can be classified into four categories: (i) a high-level language for SDN applications to define their network operation policies; (ii) a rule update process to install rules generated from those policies; (iii) a network status collection process to gather network infrastructure information; and (iv) a network status synchronization process to build a global network view using the network statuses collected by each individual controller.

(1) One of the fundamental functions of the SDN controller is to translate application specifications into packet-forwarding rules. This function advances a protocol to address communication between its application layer and control layer. Therefore, it is imperative to realize some high-level languages (e.g., C++, Java, and Python) for the development of applications between the interface and the controllers.

(2) An SDN controller is accountable for generating packet-forwarding rules as well as describing the policies perfectly and installing the rules into relevant devices. Meanwhile, the forwarding rules should be updated with policy changes. Furthermore, the controller should maintain consistency for packet-forwarding by using either the original rule set/updated rule set or by using the updated rules after the update process is completed.

(3) SDN controllers accumulate network status to provide a global view of the entire network to the application layer. The network status includes duration time, packet number, data size, and flow bandwidth. A helpful and commonly employed method for network statistics data collection is Traffic Matrix I [37]. TM controls the volume of all traffic data that passes through all sources and destinations in any network.

(4) Unauthorized control of the centralized controller can degrade controller performance. Generally, this can be overcome by maintaining a consistent global view of all controllers. Moreover, SDN applications play a significant part in ensuring application simplicity and guaranteeing network consistency.

*2.3.3. Application Layer.* As shown in Figure 3, the application layer is located at the top layer of the SDN architecture. SDN application interacts with the controller through the northbound interface to achieve an unambiguous network function in order to fulfill the network operators' requirements. They request network services or user requirements and then manipulate these services. Although there is a well-defined standardized southbound interface such as OpenFlow, there is no standard northbound interface for interactions between controllers and SDN applications. Therefore, we can say that the northbound interface is a set of software-defined APIs, not a protocol. SDN applications can provide a global network view with instantaneous status through northbound APIs. We can categorize SDN applications according to their related basic network functionality or domain including QoS, security, traffic engineering (TE), and network management. However, several SDN applications can be developed for specific use cases in a given environment.

*2.4. Applications.* SDN can modify the network configuration according to user requirements. To justify the advantages of the SDN architecture, in this survey, we present a few SDN applications.

*Wireless and Mobile.* In a wireless sensor network, SDN provides benefits such as flexibility, optimized resource allocation, and easier management. The SDN controller permits sensor nodes to support multiple applications as they have the flexibility to set any new policies or rules. In [38], SDN in ad hoc networks has been deployed to apply the concepts of abstraction to wireless ad hoc networks for smartphones. This SDN-enabled mobile infrastructure has been implemented in the Android operating system that is more secure and easier for modification and extension.

*Load Balancing (LB).* LB is an important technique for online resources management to control the data flows from different applications in order to keep the link utilization at its lowest level. Moreover, the choice of an appropriate link is very important to enhance service functionality, increase throughput, avoid network overloading, minimize cost, and reduce response time. In [39], an OpenFlow-based load balancing solution is presented. When using SDN technologies, load balancing can be integrated using the OpenFlow switch, thus avoiding the need for a separate device. Moreover, SDN allows load balancing to operate on any flow granularity.

*Network Management.* It is reported that more than 60% of network failure happens due to human configuration [40] errors and failure in order to provide an automated and comprehensive network management system. SDN provides an abstract view of the entire network, which makes network management more flexible and automated. In SDN, a network is managed from a centralized controller based on controller flow tables and flow rules that are distributed throughout the network through its interfaces, which ensures a more flexible, granular management [41].

*Network Security.* In traditional networks, firewalls or proxy servers are used to protect the physical network. SDN uses a centralized architecture to deal with network security issues. SDN supervision of flows across the entire network and monitoring of user behavior allows SDN architecture to detect and prevent damage. If attacks are detected, the SDN controller can install packet-forwarding rules in the underlying switching devices to successfully prevent the attack from entering and propagating in the network [42]. One of the problems of SDN for attack detection in the case of high network traffic is that the flow tables are not sufficient to support the high-traffic flow. Therefore, in [43], a solution in the form of a real-time security system has been proposed.

*Multimedia and QoS.* Existing network architecture is based on end-to-end data transmission, but not supported for multimedia traffic (e.g., video streaming, video conferencing, and video on demand) though in case of real-time transmission, it requires high levels of efficiency and quality with tolerable delay and error rate. According to studies by CISCO, IP video traffic will increase from 67 percent in 2014 to 80 percent by 2019 [44]. SDN provides greater QoS by effectively selecting the optimized path among all available paths. In [45], authors proposed enhancement or optimization methods for improving end-to-end multimedia QoS over SDN.

*Monitoring and Measurement.* The control application needs to monitor the link constantly in terms of latency and bandwidth to optimize data flow provisioning. SDN allows a network to perform certain monitoring operations without any additional hardware or other overheads because an SDN inherently collects information about the entire network to maintain a global network view through a logically centralized controller.

### 2.5. Challenges and Future Direction in 5G.
Cellular network technologies have experienced explosive evolution during the last decade. Moreover, the number of mobile devices and the data traffic are increasing exponentially because network applications are extending from the traditional hardware-based to real-time communication in social networks, e-commerce, and entertainment. However, hardware-based cellular systems depend on insecure and inflexible network architectures which generally take a typical 10-year for a new generation of wireless networks to be standardized and deployed. Nowadays, the most emerging cellular network system is 5G [46]. But 5G has some challenges to think about [47]. In particular, the requirements for the 5G network system are high data rates (targeting 1 Gbps experienced users everywhere), ultra high capacity should be 1000-fold capacity/km$^2$, cost, a massive number of connections, and E2E latency should be less than 1 ms over the RAN. To facilitate these challenges of current network architectures, the most important need is to shift the design of current architectures for the next generation wireless networks. Moreover, the complementary concept of SDN, NFV has been presented to effectively separate the control functionalities from the hardware by simply decoupling the forwarding plane from the control plane. These functionalities will ensure the required flexibilities and adaptability of the ever-changing cellular network architectures with the introduction of the concept of SDN for 5G.

Though we presented the advantages of SDN for 5G, SDN was confronted with some challenges. First of all, security is a more challenging task that needs to be available everywhere within the SDN architecture because of (i) architecture and its controller, applications, devices, channels (TLS with plain text) and flow table, (ii) connected resources, (iii) services (to protect availability), and (iv) information. Furthermore, a reliable and balance controller is still out of scope because of lacking of robust and reliable framework policy. The framework policy should be very simple to maintain and implement, secure, and cost effective. An integration of SDN with NFV can be a new category for security deployment by decoupling control plane from forwarding plane or switching devices. In addition to security, link and controller availability, reliability, flexibility, controllers, and applications compatibility are considerable concerns. A centralized controller is not as fast as it is supposed to be though it can recover itself through a backup flows checking [48].

Operational, maintenance, and fixed costs are also another challenging topics for the deployment of 5G. The expenses can increase to reduce system blockage and maintain the availability though integration of SDN and NFV can reduce expenses. As a fully automated system with a centralized controller, SDN offers reduced human control and error free and fast configuration [48]. Despite these challenges, some remaining implementation affairs need to acknowledge such as flow tables and their large number of flow entries, flow level programming and controller programming, flow instructions, and actions.

NFV and SDN are independent and complementary to each other. But they can provide an open environment to fasten the innovation and can be easily integrated with new services and infrastructure like controlled and automated network resources. This combination can easily manage resources using its centralized controller. Packet-forwarding or processing is performed by NFV, while the controller can control or update flow tables according to the needs of users or applications at any level. Here, NFV is responsible for creating or processing flow rules, and SDN is responsible for the management of the said rules. Integration of SDN and NFV will be a promising technology for 5G.

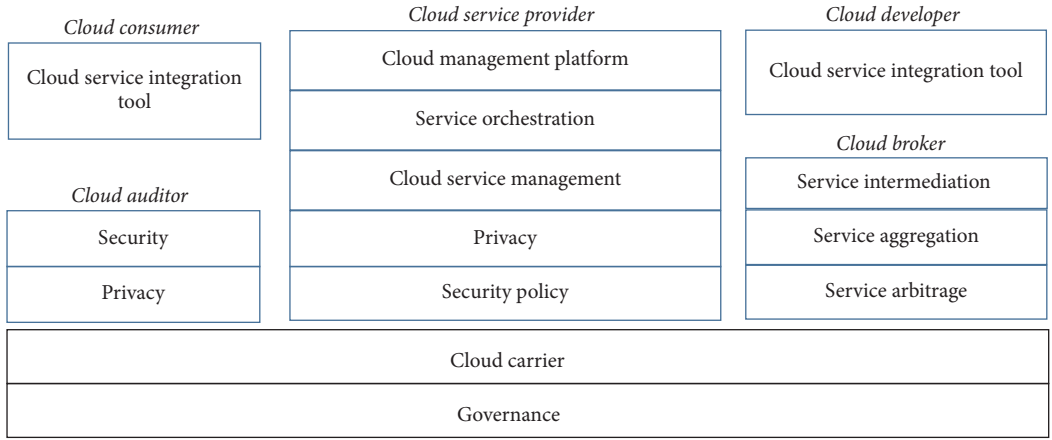

FIGURE 5: Cloud computing interface architecture.

## 3. Cloud Computing

*3.1. Cloud Computing for 5G Network.* The main characteristics of 5G include high speed, low latency, and high capacity to support various real-time multimedia applications. 5G is being developed as a smart wireless network architecture using new models such as SDN or NFV for multidimensional massive data processing [49]. Network virtualization is a new concept that can create a big challenge for next generation networking based on IP networks, the Internet, and wireless technology. Virtualization creates connections between the communication and the computing domains. Service-Oriented Architecture (SOA) will be the main factor of network-as-a-service (NaaS), which is enabled by the convergence of networking and cloud computing. Network virtualization architecture with SOA has attracted wide interest from both the academia and the industry. However, some issues related to user requirements of QoS and QoE remain. There are more and more services considering cloud computing as the core backbone technology for service deployment and network implementation due to the scalability and flexibility. Cloud service is an important technology for the future, as it can reduce costs for service provider and customer by efficient resource allocation. Cloud computing becomes one of important reference architectures for 5G network due to the high data rate, high mobility, and centralization management services. It can be operated without direct installation consumers' systems. Cloud computing has been considered increasingly by both the academia and the industry. Due to development of technology and business trend of mobile service, the number of mobile devices is increasing more and more. The world is switching to compact devices with limited computing power, cloud computing will be the future of consumer technology. To provide a common protocol of management and operation for cloud service deployment, a complete and precise standard architecture is required. The researches on cloud computing issues are being carried out based on many cloud computing projects. They are attempting to standardize solutions related to architecture, operation, authentication, service, and cooperation integration of 5G network. The general interface architecture of cloud computing from proposals is selected from Figure 5. Based on the architecture and network function, cloud computing research can be classified by topology framework, architecture and service.

For topology frameworks, "CloudAudit: Automated Audit, Assertion, Assessment, and assurance" [50] was officially launched in 2010. It offers a cloud computing service architecture based on open, extensible, and secure interface and methodology. Cloud Standards Customer Council [51] deals with standards, security, and interoperability issues. Cloud Storage Initiative [52] discussed storage issues in cloud services by considering the adoption of cloud storage as a new delivery model. OASIS Identity in the cloud (IDCloud) [53] works on open standards for identity deployment, provisioning, and management in cloud computing. OpenStack [54] provided an open-source software API for private clouds. "Cloud Computing Interoperability Forum" [55], Open cloud Consortium [56] focused on cloud integration framework. MCC [57] proposed a new framework for 5G cloud computing by enhancing the traditional MCC architecture to satisfy the requirement of QoE in emotion-aware applications. It has three main components: mobile terminal, local cloudlet, and remote cloud. The proposed system can support the latest technological advances of 5G with computation-intensive affective computing, big data analysis, resource cognition-based emotion-aware feedback, and optimization of resource allocation under dynamic traffic load.

For service architecture, open cloud frameworks such as platform-as-a-service (PaaS) and infrastructure-as-a-service (IaaS) have been proposed by CloudFoundry [58] and DeltaCloud [52]. Open cloud computing Interface [59] provided interfaces for cloud resource management, including computing, storage, and bandwidth. Cloud Security Alliance [60] and Distributed Management Task Force (DMTF) [61] deal with cloud computing security. Open Data Center Alliance [62] develops usage models for cloud vendors with long-term data centers. The proposed cloud computing architecture in [63] uses an actor-based structure. It comprises six major actors: cloud consumer, cloud provider, cloud developer, cloud broker, cloud auditor, and cloud carrier. The actors have their own activities, requirements, and responsibilities. The associated cloud services are classified into four different groups: IaaS, PaaS, software as a service (SaaS), and anything as a service (XaaS).

For standard reference architecture, Standards Acceleration to Jumpstart Adoption of cloud computing [64] and The Open Group cloud Work Group [60] are examples of the use case creation of cloud computing standards. TM Forum cloud Services Initiative [65] suggested approaches to increase cloud computing adoption on different network service. CloudCommons [66] evaluated cloud service business performance based on the provided set of service measurement index (SMI). The proposed architecture focused mainly on commercial models, service functions, measurement of service-users preferences, and satisfaction indexes. Trusted Cloud Initiative (TCI) reference architecture was proposed by Cloud Security Alliance (CSA) in 2011 [67]. TCI uses four frameworks to define its security polity, namely, the Sherwood Business Security Architecture (SABSA), Information Technology Infrastructure (ITIL), the Open Group Architecture Framework (TOGAF), and Jericho. The proposed architecture includes methodology and supporting tools for security configuration, enterprise architecture, business plan, and risk management. The business requirements are based on different standards of control matrixes, payment, authentication, planning, design, and service development. The architecture is complex from the viewpoints of implementation and deployment because it combines several different frameworks, thus requiring developers to understand all frameworks. The standard of cloud computing is still an open issue because different industries have different precise definitions based on their own architectures [63]. However, the National Institute of Standard and Technology (NIST) [68] and IBM [69] are two typical cloud computing architectures that have been applied as references by both the industry and the academia. NIST introduced research on cloud computing architecture in September 2011 by suggesting a reference architecture including the main elements of cloud computing. The proposed architecture provides categories of functions, activities, and classification methods based on a tree structure. It describes the general concepts of technical functions and business models. The service management functions in the architecture require background knowledge. The architecture should have additional explanation and operational description for nonbackground users [63]. The detailed architecture proposed by IBM's research team is called Cloud Computing Reference Architecture (CCRA), and it is based on customer's demands of IBM's cloud products and services. Experience and research pertaining to cloud services were applied to devise a full cloud computing architecture. Compared with NIST, IBM CCRA has important advantages in terms of operation and management. System performance and scalability are considered based on the customer's cloud computing environment. Another strong point is the support system. IBM provides specific development and management tools to help customers deploy and manage their cloud services.

Cloud computing creates a new paradigm in networking technology with the concept of computing resource sharing. It can provide ubiquitous on-demand access with high flexibility, cost efficiency, and centralized management. Cloud computing has attracted considerable attention from and has had an impact on the ICT community. An increasing number of critical applications and services now support cloud computing architecture. Cloud computing is a promising architecture for future-generation networks. The distributed, dynamic, and heterogeneous characteristics of resource management are the main difference between cloud computing and the traditional service model, where the available architecture of the traditional network cannot adapt to new features. The resources used in cloud computing have different features. Hence, with the static QoS index strategy, system performance is not efficient.

*3.2. Challenging Issues and Future Directions in 5G.* Cloud computing creates a new paradigm for networking technology with the concept of computing sharing and distributing resource. It can provide ubiquitous on-demand access with high flexibility, cost efficiency, and centralized management. Cloud computing has attracted considerable attention from many researchers and organization. With important contributions on architecture, cloud computing has had an impact on the ICT community. There are more and more critical applications and services which support cloud computing architecture. It will be promising architecture for future-generation of networking. Compare with traditional network and service architecture, cloud computing has advantages on distributed, dynamic, and heterogeneous characteristics of resource management. With the development of technology on semiconductor and human on demand, the traditional network architecture shows some limitations on mobility functions which cannot follow new features especially the static QoS index strategy.

With the advantage of higher capacity and powerful accessibility, 5G will be an enhanced technology with full on-demand mobile applications and services. In addition, it gains from the development of other services such as social networks, wearable devices, IoT, and cloud computing. Traditional network applications will be more human-centric on demand. A QoS model that can be configured dynamically based on the description of the required resource QoS is introduced in 5G by using three models: series, parallel, and hybrid [70]. The service architecture of cloud computing generally is categorized into three classes: SaaS, IaaS, and PaaS. The layered structure of cloud computing services is shown in Figure 6. SaaS includes applications such as Google Apps, Salesforce, and Microsoft Office 365. IaaS includes applications such as Amazon cloud Formation, Google Compute Engine, and Rackspace cloud. This service model defines the service for users of servers and storage. IaaS supports users with an interface for virtual management and storage. PaaS includes applications such as Google App Engine, Microsoft Azure, and Amazon Elastic Beanstalk. Platform-as-a-Service provides access APIs, programming languages, middleware, and framework, which can be developed according to a user's applications without installing or configuring the operation environment. One of the most daunting challenges faced when working with PaaS is ensuring compatibility because there are no common features, API database type tools, and software architectures across various PaaS architectures.

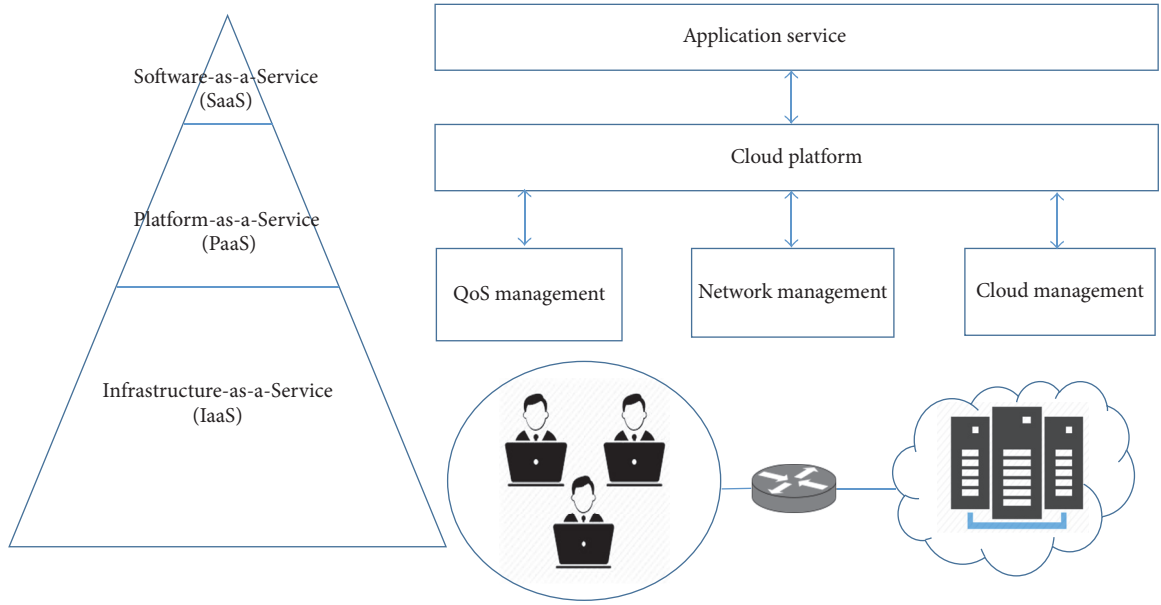

FIGURE 6: Cloud computing service layers.

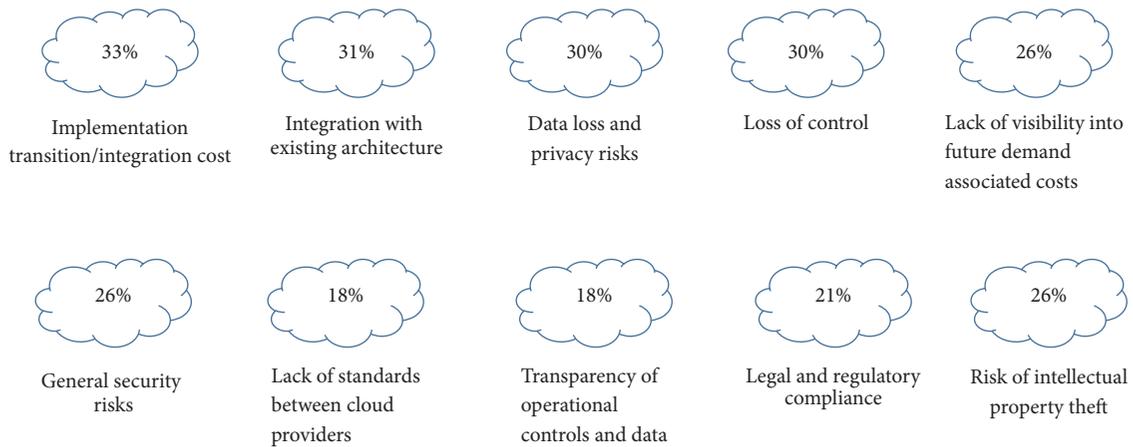

FIGURE 7: Challenges of loud computing service shifting.

Besides the advantages of cloud computing service from traditional network architecture, there also remain some certain challenges during the cloud shifting. Figure 7 shows the considerations of adopting cloud computing from implementation challenge survey by KPMG [71]. For 5G network, the challenge issues of cloud computing are considered as Security and Privacy, Quality of service, access time and Accessibility, Data access control, and transition to the cloud. About Security and Privacy issue, they are the most concerns for service providers when moving their data to the cloud. Although Security and Privacy issue in most cloud architectures is generally designed with high reliable and proficient model, it must show a fully secure scenario with different levels and strategies for customer's trust. Secondly, the Quality of service will be considered strongly before moving from traditional network architecture to cloud architecture. The cloud customer needs assurance that the business's data will be safe and available and reliable at all times. The performance of cloud infrastructure can be affected by the load, environment, number of users, and connection link technology. There should be some backup mechanisms to guarantee the data access. The research issues for access time and accessibility are related to infrastructure of cloud architecture. With data access control issue, there are some questions on the reliability of system such as backup strategies, storage structure, and security of data access. The transition from traditional network to cloud is important time of cloud customer. The first step in transitioning to the cloud is being able to identify the challenges and working conditions with cloud provider to navigate the barriers of cloud business model.

TABLE 2: Leading organizations, institutions, and forums involved in IoT standardization.

| International organizations | Regional and national organizations | Global standards collaboration | Forums | Clusters |
|---|---|---|---|---|
| (i) ITU-T [75–77]<br>(ii) IEEE [74, 78–81]<br>(iii) ISO [82]<br>(iv) IEC<br>(v) ISO/IEC JTC1 [83, 84] | (i) ETSI [85, 86]<br>(ii) CEN<br>(iii) CCSA<br>(iv) ARIB<br>(v) TIA<br>(vi) TTA<br>(vii) TTC<br>(viii) GISFI | (i) MSTF | (i) oneM2M [87]<br>(ii) W3C<br>(iii) 3GPP<br>(iv) NFC<br>(v) ECMA<br>(vi) IoT Forum<br>(vii) Ipv6 Forum | (i) IERC |

## 4. Internet of Things (IoT) for 5G

### 4.1. IoT Definition.
IoT is a dynamic network of connected devices. The idea is to connect not only things but also people any time, any place, with anything and anyone, and so on. The definition of IoT has crossed the boundaries of traditional network. The International Telecommunications Union (ITU) has codified the concept of IoT [72] as the following definition:

> "IoT is a global infrastructure for the information society, enabling advanced services by interconnecting (physical and virtual) things based on existing and evolving interoperable information and communication technologies".

However, IoT has become hugely popular over the last decade. The dimensions and the scopes of IoT can be any Thing, any Place, any Time, any Body, and so on. Consequently, standardization is being demanded to establish interoperability among things with a view of transforming the world into a global village. The standardization efforts undertaken by different organizations and institutes are explained below.

### 4.2. Standardization Effort.
Standards control any system to operate under fixed rules and regulation. Interoperability among the disciplines of any reference system depends on standards. Worldwide, numerous standardization authorities have initiated the creation of relevant standards during the last decade. Nevertheless, these efforts have had no impact in terms of unifying the standards into a single framework because IoT has become the storehouse of anything. The list of different organizations, institutions, and groups engaged in IoT standardization is given in Table 2 [73]. A few persuasive IoT-related standards by IEEE are listed in Table 3 [74].

### 4.3. IoT Architecture.
The future is approaching a new paradigm of networks with huge numbers of devices. The idea of 5G (beyond 4G) refers to networks with improved QoS, huge capacity, enhanced data rate, and, overall, a feasible architecture to sustain the aforementioned features. The influential parts of 5G networks include D2D communication, which can be interpreted as the idea of IoT. IoT comprises the technologies of smart sensors, RFID, machine-to-machine (M2M), IP, communication systems, and so on. This part of the paper focuses on the different emerging IoT architectures suitable for future-generation 5G networks.

IoT architecture has evolved with the evolution of the Internet. The first phase of IoT evolution entailed communication among several computers through a computer network. However, the World Wide Web (WWW) was launched in 1991 to connect all computers worldwide [89, 90]. Further technological advances have connected the users of various types of electronic devices with computers under the same platform by connecting to the cloud network [91]. Finally, the idea of IoT was conceived to give shape to the world by connecting everything. IoT is the network that can adopt and connect anything that anyone can imagine [92].

IoT architectures can be classified into several types because it is absolutely difficult to merge the architectures proposed for various IoT applications into a single model [93]. A scheme for classifying IoT architectures is shown in Figure 8. Several authors have proposed three-layer-based simple IoT architectures that comprise an application layer, a network layer, and perception layer [94, 95]. Middleware-based IoT architectures consist of a greater number of layers, including the coordinate layer next to the middleware layer [96, 97]. In addition, the perception layer has a shared option for combining other edge technology and the access layer [98, 99]. Service-oriented architecture (SOA) has different layers, unlike middleware-based architecture. It has five layers, namely, the objects layer, object abstraction layer, service management layer, service composition layer, and application layer [100, 101]. However, common IoT networks have an architecture comprising five fundamental layers, namely, the objects layer, object abstraction layer, service management layer, application layer, and management layer.

The first and foremost layer is the *objects layer*, which is similar to the perception layer that embodies physical devices, and an IoT architecture might contain heterogeneous devices in the network. Next is the *object abstraction layer*, which is used for the conveying the data generated by the devices [102]. Various technologies are used for data transfer, for instance, the 5G network uses RFID, WiFi, Bluetooth, UWB, and ZigBee. Cloud computing technology, too, is deployed in this layer. Then, there is the *service management layer*, and it is responsible for application programmer management, which entails ensuring compatibility with any hardware platform by processing the generated data.

TABLE 3: Standardization efforts for IoT by different groups of IEEE.

| Group | Title of the Standardization Group |
| --- | --- |
| IEEE 802.11-2012 | IEEE Standard for Information Technology–Telecommunications and information exchange between systems–Local and metropolitan area networks–Specific requirements Part 11: Wireless LAN Medium Access Control (MAC) and Physical Layer (PHY) specifications Amendment 10: Mesh Networking |
| IEEE 802.15.4-2011 | IEEE Standard for Local and metropolitan area networks–Part 15.4: Low-Rate Wireless Personal Area Networks (LR-WPANs) |
| IEEE 802.15.4g-2012 | IEEE Standard for Local and metropolitan area networks–Part 15.4: Low-Rate Wireless Personal Area Networks (LR-WPANs) Amendment 3: Physical Layer (PHY) Specifications for Low-Data-Rate, Wireless, Smart Metering Utility Networks |
| IEEE 802.15.7-2011 | IEEE Standard for Local and Metropolitan Area Networks–Part 15.7: Short-Range Wireless Optical Communication Using Visible Light |
| IEEE 802.16-2012 | IEEE Standard for Air Interface for Broadband Wireless Access Systems |
| IEEE 802.16p-2012 | IEEE Standard for Wireless MAN-Advanced Air Interface for Broadband Wireless Access Systems – Amendment: Enhancements to Support Machine-to-Machine Applications |
| IEEE 802.16.1b-2012 | IEEE Standard for Wireless MAN-Advanced Air Interface for Broadband Wireless Access Systems – Amendment: Enhancements to Support Machine-to-Machine Applications |
| IEEE 1609.11-2010 | IEEE Standard for Wireless Access in Vehicular Environments (WAVE)–Over-the-Air Electronic Payment Data Exchange Protocol for Intelligent Transportation Systems (ITS) |
| IEEE 1888-2011 | IEEE Standard for Ubiquitous Green Community Control Network Protocol |
| IEEE 1901-2010 | IEEE Standard for Broadband over Power Line Networks: Medium Access Control and Physical Layer Specifications |
| IEEE 1905.1-2013 | IEEE Draft Standard for a Convergent Digital Home Network for Heterogeneous Technologies |
| IEEE 11073-10103-2013 | IEEE Standard for Health informatics – Point-of-care medical device communication – Nomenclature – Implantable device, cardiac |

The *application layer* affords customers with services as requested. This layer includes different types of services such as smart city [103], smart wearable device [104], smart vehicle [105], smart home [106], smart healthcare [107], and industrial automation [108].

The self-characteristics of IoT should cope with the emerging future-generation 5G networks. The basic infrastructure of IoT has the characteristics of heterogeneous devices, resource-constrained, flexible infrastructure, dynamic network, distributed network, ultra-large-scale network, large number of events, spontaneous interaction, location awareness, and intelligence [109].

However, the idea of the future-generation 5G network possesses the characteristics of IoT networks. The simplified architecture of the 5G network is shown in Figure 9 to demonstrate that the emerging IoT architecture can deal with it. It is assumed that the 5G network might connect 50 billion devices to the cloud by 2020. The 5G network will cause a 10–100x increase in the number of devices, 10–100x increase in data rate, 1000x increase in data volume, 10x increase in device battery life, and 5x decrease in latency. The 5G network embraces some important technologies such as radio access, MIMO, mobility management, interference management, and massive spectrum to achieve compatibility with the IoT networks included in the METIS project. To handle different issues with these technologies, some mechanisms have been proposed in the METIS project [110]. D2D communications is one of the proposed methods that helps maintain an ultra-large-scale network using flexible infrastructure. However, massive machine communication (MMC), another solution, is the base of IoT in terms of interconnecting a huge number of devices across different smart technologies. Furthermore, moving network (MN), ultra-dense network (UDN), and ultra-reliable network (URN) are some other proposed solutions for mobility management, interference mitigation, capacity achievement, and so on.

Several studies on different types of services for IoT have been and are being carried out. A few vital architectures of the enabling technologies for the future-generation 5G network have been surveyed. M2M-based communication architecture for cognitive radio network has been demonstrated for IoT [111–113]. The architecture of the M2M network in IoT is shown in Figure 10. The relationship between the infrastructure layer and the application layer is maintained by the network layer, which is mainly a communication network. The infrastructure layer includes the M2M devices and gateway, while the application layer comprises the users, management interface, and the M2M server.

The M2M server is the core of the architecture, and it integrates the overall system for the required services such as traffic management, smart healthcare services, and so forth. However, the object database (DB) also sends user information preserved in the DB by users in the form of SMS, email, video, and so on. The IMS server is connected to the M2M server, which is situated in the network layer. The GPRS server and gateway are the main components of the network layer that help the IMS server by collecting vehicle location in the system. The application layer and the network layer are connected through the Internet, while the infrastructure layer and the network layer are linked together by the gateway or some protocol. M2M devices update the M2M server through the network layer by using the information of M2M devices (i.e., body sensors, smart devices, other devices). The user interface provides access to the user and manages user data. The network layer, which consists of a WPAN/WLAN network and an IoT gateway, connects the user interface with the management interface. The management interface monitors user data, takes initiatives as necessary, and informs the respective body about the situation [114–116].

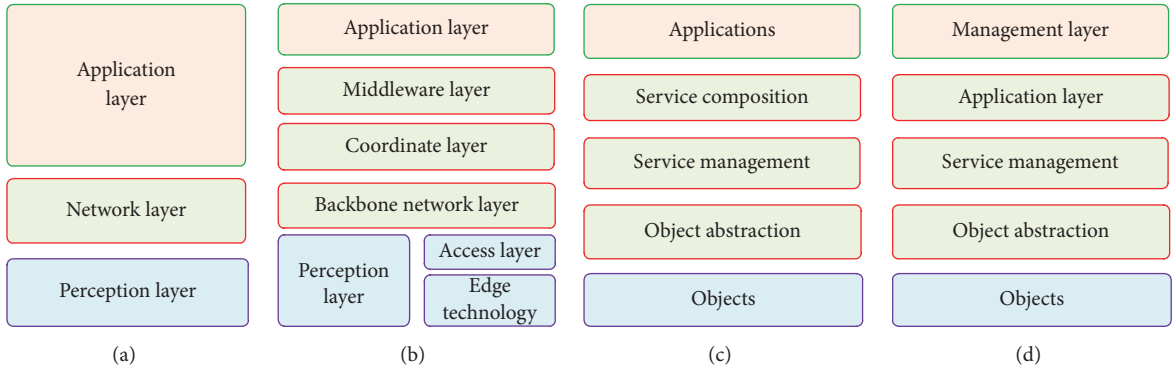

FIGURE 8: Categorized IoT architectures. (a) Three-layer-based simple architecture, (b) middleware-layer-based architecture, (c) service-oriented architecture (SOA) for IoT, and (d) five-layer architecture (adopted from [16]).

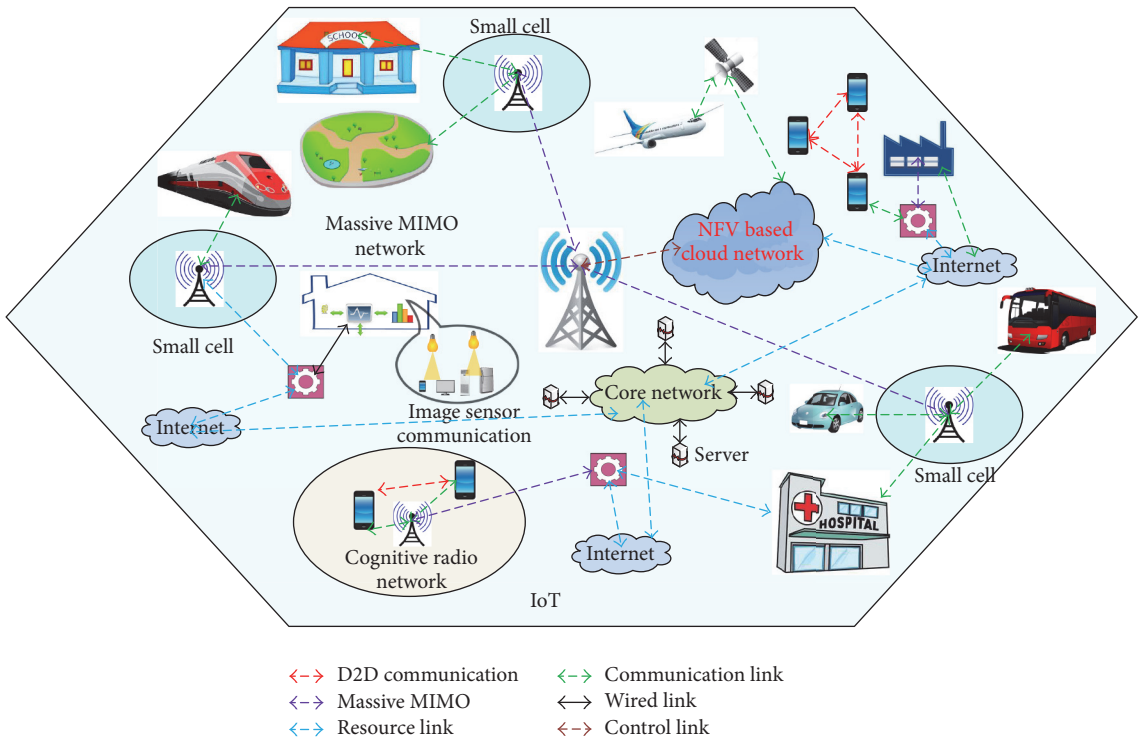

FIGURE 9: Simplified future-generation cellular 5G networks.

*4.4. Applications.* IoT has become the source of many applications owing to its incredible potential and has given rise to numerous new application fields. It has brought revolutionary changes to our everyday life. It has connected everything, everyplace, and everybody to an inseparable framework, which has shrunken burdens many times over multiple associated systems. However, the applications of IoT are indescribable because IoT can contribute to almost every sector. Hence, the impact of IoT has surpassed our social and economic life and has entered our personal life. Figure 11 shows a summary of the numerous IoT applications. Note that it is quite impossible to outline all the applications in a frame. We describe a few of the influential IoT applications below.

*Smart Home.* The idea of automation in home management and surveillance has brought personal life under the supervision of the IoT platform. Home appliances can be controlled from a remote place using IoT technology [117, 118]. Furthermore, human-machine interaction for the smart home environment is a new inclusion in IoT [92, 119, 120]. Wearable devices can be helpful in human-machine interaction and for home monitoring [121]. The environment of a smart home can be controlled effectively using the smart home technology in IoT [106].

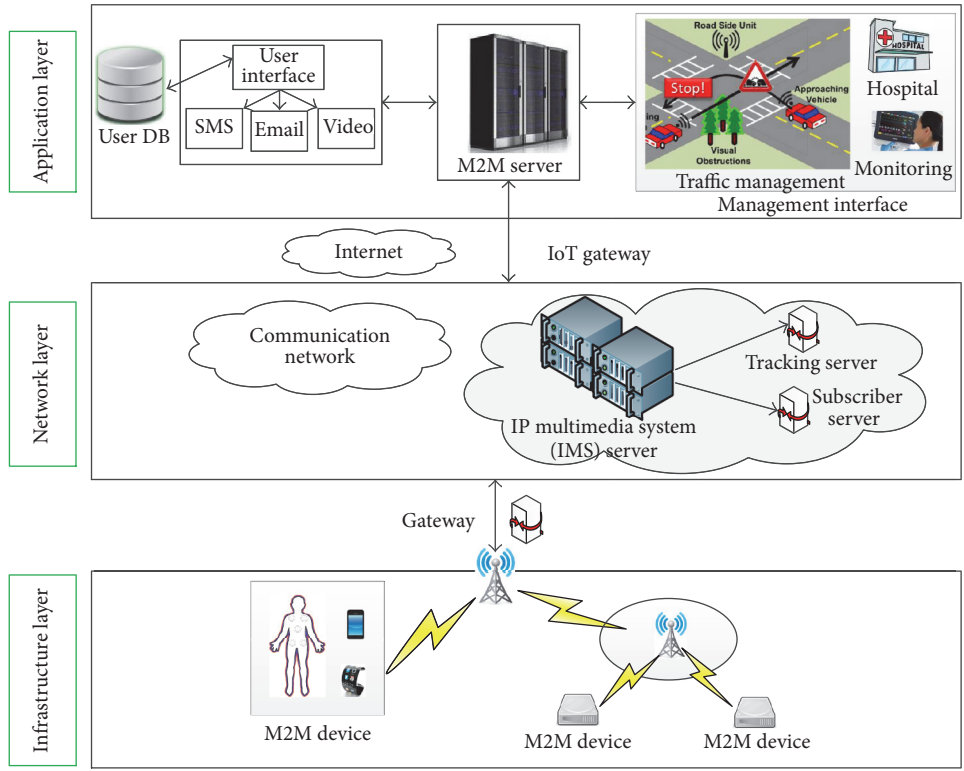

FIGURE 10: Generalized M2M network architecture in IoT.

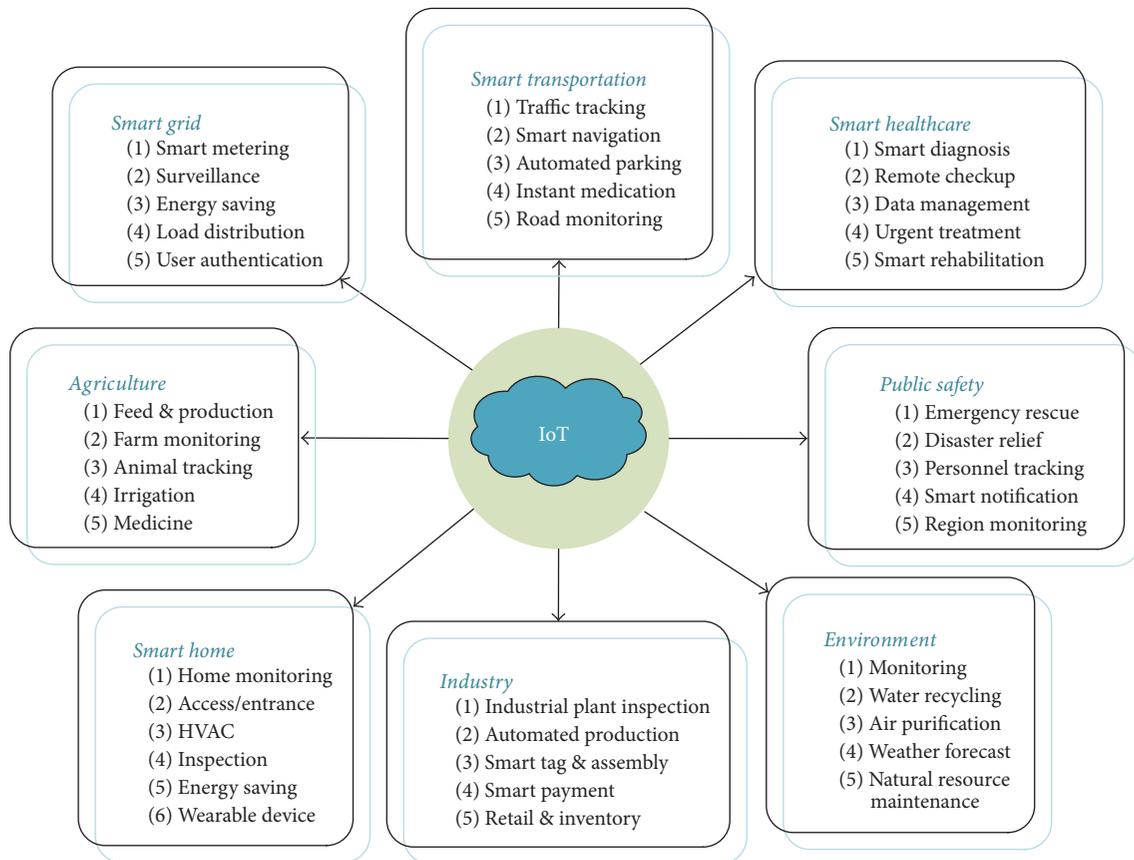

FIGURE 11: Prospective applications of IoT.

*Smart Grid.* The energy consumption and distribution of a power plant and grid can be controlled by connecting them to IoT. The use of communication technology in the smart grid system establishes a link between user and system, helps broadcast urgent information to the customer, facilitates automatic device control, and so on [122]. However, the smart grid is closely related to the smart home system and can help reduce the energy consumed by such a home [123, 124].

*Smart Transportation.* IoT has brought of the idea of smart city to ameliorate the provision of basic services. Vehicle tracking is one of the promising applications of IoT that can lessen traffic congestion, enhance safety, schedule traffic time and smart vehicle parking, send information to travelers, and so on [125–127]. Cloud computing, the catalyst for IoT, has added a lot to the smart management and automation of smart parking systems [128].

*Smart Healthcare.* IoT has flourished smart healthcare systems by interconnecting various devices [129]. Smart healthcare systems cover healthcare records [130], remote prescription [131], patient observation [132], urgent treatment [133], smart diagnosis [16], and so on. Smart rehabilitation can be a part of a smart healthcare system to support the elderly and the disabled to get medication [134].

*Public Safety.* Safety and security are the most essential issues in a smart city. IoT contributes to emergency alarm systems, weather forecasts, disaster management, and emergency evacuation, and so on [135–137]. Moreover, intelligent algorithms and camera based system have progressed the safety system in a distinguished way [138].

*Agriculture.* Agriculture includes planting, farming, breeding, and animal rearing, and it has come under the purview of IoT lately [139]. The monitoring of farms, tracking of animals, and irrigation are the main aspects of IoT for agriculture [140, 141]. Moreover, feeding, vaccination, medication, rearing, and so on are vital applications of IoT in the agriculture sector [142].

*Industry.* The integration of different enterprises in the industry has brought about a radical change in this sector [143]. Automated monitoring systems for CO2, poisonous gases, and other gases can be among the great applications of IoT [144]. In addition, RFID and wireless sensors can be potential fields for automation in industries [145].

*Environment.* A physical environment may be arranged with smart devices, and the automation of homes, industries, traffic, and healthcare can be connected to physical entities to build a better world [146]. Disaster management, weather forecasting, and emergency alarm service are among some of the essential applications of IoT [147]. However, the concept of green IoT technology has been established to create a smarter environment [148, 149].

### 4.5. Challenges and Future Direction

*4.5.1. Challenges.* The definition of IoT has the significance of the challenges for 5G. The pivotal features of IoT such as heterogeneity, secure communications, system protocols, and so forth have commenced different challenges for IoT. Some of the key challenges are described below briefly.

*Large Scale Storage.* The property of heterogeneity creates a huge demand for storage of data. Besides, various types of data need to be categorized for the simplification of computation, data generation, and processing, which enhance the necessity of storage size of data.

*Computation.* One of the critical challenges of IoT, which emerged as one pivotal issue, is computation. The integration of heterogeneous device and functional variance of the devices has aggrandized the computation problem. The architecture of IoT demands a reliable and scalable computational methodology.

*Ubiquitous Protocol Design.* The architecture of IoT initializes a common platform for the devices with the different working mechanism. The D2D communication, an essential feature for IoT, produces a challenge for the IoT of 5G to build a pervasive protocol for connecting the heterogeneous devices. Every device connecting to each other should maintain a common protocol to make computation process simpler and to bring the characteristics of scalability.

*Security and Privacy.* Security, as well as privacy, has transformed into a major challenge for IoT. As cloud computing is used in IoT for storage of data, security has become one of the primary concern for the virtual storage. Moreover, lacking personal privacy between numerous devices has made the problem more critical because the establishment of personal privacy at each layer of the IoT architecture requires computing power constraints [128].

*Reliability.* Reliability arises as one of the major concerns recently. Due to the connectivity of everything, reliability in certain sectors can be defined as the most serious challenge. Public health such as emergency operation, critical treatment for diseases, smart transportation, and smart home are some of those types areas of IoT in which reliability plays a vital role.

*Performance.* Performance of IoT varies according to the several activities of the different layer of IoT. Particular functions in IoT need highly assured performance and QoS. Traffic mobility, real-time connections, and emergency services are some of them that have inaugurated the challenge of performance in IoT.

*4.5.2. Future Direction.* The introduction of cloud computing, big data, SDN, and so forth in the IoT area has initialized a new era of research. The standardization effort of IoT has also imposed an influence on the future research of IoT. The methodology of integration of several devices of the different mechanism is one of the significant issues for future studies. Several system architectures have been proposed to make IoT architecture viable, but all of the systems have been proposed on the different platform. It is a matter of concern how all of the systems could work on the same platform with massive computational performance and energy efficiency. Security, privacy, and reliability are the next candidates for the common platform. Consequently, the integration architecture of big data, cloud computing, and SDN for IoT would be the most important issue for the future direction of research.

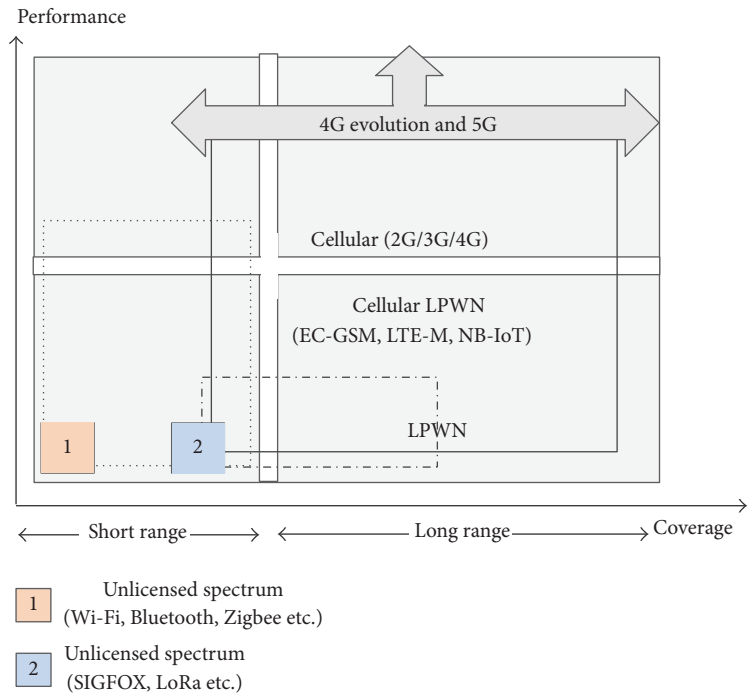

FIGURE 12: Technologies addressing different M2M segments [17].

However, several other aspects of IoT can also be important parts of future research direction. Intelligent system for real-time data collection and processing in the IoT architecture could be one of the important research directions. Establishment of individual small social networks to work for different devices and combination of these networks to create a single platform for IoT could be another future direction for the research area. Besides, scalability, reliability, and flexibility are some other issues for future direction.

## 5. Mobile Access Networks for 5G

*5.1. M2M (Machine-to-Machine) Communication.* Legacy cellular networks have been developed to support high data rate and reliable communication. M2M environments are, however, very different from cellular networks, because low data rates and long latencies are desirable. The basic purpose of M2M communication is to transmit sensed data of small size with loose time constraints. To meet the characteristics of M2M communication, there are two categories of Radio Access Technologies (RATs) according to spectrum resources: cellular IoT and lower-powered wide-area network (LPWN). A classification of cellular IoT technologies is shown in Figure 12.

Cellular IoT involves modifying the legacy cellular network to accommodate IoT communication using licensed bands. The third-generation project partnership (3GPP) standardized long-term evolution machine-to-machine (LTE-M), which optimized the IoT protocol over the LTE system since Release 12 [150]. LTE-M reuses LTE PHY channels. LTE-M includes coverage enhancement, cost reduction, and improved battery life. Furthermore, it is able to cooperate within the legacy LTE network. However, it has a limitation in fulfilling all requirements of IoT communication because the nature of the LTE system is not suited for low data rates and long-range communication.

Therefore, 3GPP is currently studying and standardizing a narrow band radio interface called narrow band (NB) IoT. This technology started as a clean state standard to fulfill the requirements of IoT environments. NB-IoT reuses LTE core networks; thus, rapid deployment is possible in the market with only software modifications. In addition, NB-IoT supports various operation modes including in-band, guard-band, and standalone. NB-IoT requires only a narrowband carrier of 200 kHz with frequency division multiple access (FDMA) in uplink and orthogonal frequency division multiple access (OFDMA) in the downlink for 200,000 connections [88].

One of the benefits of using unlicensed spectrum is the ability to deploy a new service regardless of whether the service provider is an Internet service provider (ISP). Such solutions include SIGFOX [151] and Long Range (LoRa) [152]. Table 4 shows a comparison of RATs for IoT in terms of transmission range, bandwidth, data rate, battery life, and availability. SIGFOX is growing rapidly in Europe. The main target of SIGFOX is ultra-low-end sensor systems with limited throughput demands (12 bytes per 1.6 sec frame, 140 transmissions per day). SIGFOX uses the 100 Hz ultra-narrow band and basic modulation of binary phase shift keying (BPSK). It has a unique feature that all data from devices are transmitted to the SIGFOX cloud through an SIGFOX gateway and any service provider can access the data using the SIGFOX open application programming interface (API). Thus, all communication services depend highly on the SIGFOX itself because of this architecture.

TABLE 4: Comparison of various RATs for IoT connectivity [88].

| RAT | SIGFOX | LoRa | C-IoT | NB LTE-M R13 | LTE-M R12/13 | 5G |
|---|---|---|---|---|---|---|
| Range | <13 km, 160 dB | <11 km, 157 dB | <15 km, 164 dB | <15 km, 164 dB | <11 km, 156 dB | <15 km, 164 dB |
| Spectrum Bandwidth | Unlicensed 900 MHz 100 Hz | Unlicensed 900 MHz, <500 kHz | Licensed 7–900 MHz <200 kHz or dedicated | Licensed 7–900 MHz <200 kHz or shared | Licensed 7–900 MHz <1.4 MHz or shared | Licensed 7–900 MHz shared |
| Data rate | <100 bps | <10 kbps | <50 kbps | <150 kbps | <1 Mbps | <1 Mbps |
| Battery life | >10 years | >10 years | >10 years | >10 years | >10 years | >10 years |
| Availability | Today | Today | 2016 | 2016 | 2016 | Beyond 2020 |

LoRa is being standardized by the LoRa alliance since 2015. It has a communication range of 10 miles and low power consumption, resulting in a maximum battery life of 10 years. The LoRa network architecture uses general frequency shift keying (GFSK) or LoRa modulation with the star-of-stars topology and a LoRa gateway, which relays data between an end device and the core network. Each device can establish multiple connections with two or more gateways. The main difference from SIGFOX is that LoRa follows the open ecosystem policy of the LoRa alliance. LoRa uses a narrow band of 125 kHz and transmits a payload of 50 bytes with chirp spread spectrum, which is similar to CDMA. LoRa also provides adaptive data rate (ADR) to improve power management and data rate simultaneously by dynamically adjusting the data rate and transmission power based on the analytic result of packet error rate, signal-to-noise ratio (SNR), and received signal strength indicator (RSSI).

*5.2. Device-to-Device (D2D) Communication.* Device-to-device (D2D) communication in cellular networks is an emerging technology that enables direct communication between user equipment (UE) with little or no help from the infrastructure such as eNodeB or core networks. D2D communication provides several advantages in terms of spectrum efficiency, power management, coverage expansion, and capacity improvement by reusing radio resources and allowing network functionalities to devices. Furthermore, D2D communication enables new services such as public safety services, location-based commercial proximity services, and traffic offloading [153]. Owing to these benefits, D2D communication is considered one of the key techniques. D2D communication can be classified into three types based on intervention from infrastructure with network control: autonomous D2D, network-assisted D2D, and network-controlled D2D.

In autonomous D2D, devices in the network work in a fully distributed manner to communicate and establish links with each other. It is similar to ad hoc or peer-to-peer (P2P) networking. Each device or cluster head handles all network functionalities, similar to self-organizing networks. Thus, this mode is suitable for disaster networks or public safety services because devices can communicate without any infrastructure. In the case of network-assisted D2D, the infrastructure supports some network functions including link management, synchronization, and security. The devices in the network basically construct a self-organizing network and retain control over D2D communication. The infrastructure mediates network nodes to improve network efficiency by reducing the control signaling overhead. In the case of network-controlled D2D, the infrastructure strongly controls the network from radio resource management to data communication. When the network is fully centralized, all D2D devices are allowed only for data communication. It is close to the existing cellular mode.

We can categorize D2D communication types into in-band D2D and out-band D2D as well, regarding spectrum resources [154]. In in-band communication, cellular and D2D devices share the same spectrum band by reusing radio resources (underlay) or using dedicated resources (overlay). The advantage of this type of communication is that the infrastructure can have a high-level of control over the cellular spectrum, but there is additional interference from D2D communication to cellular communication, which requires an additional computation procedure for resource allocation, resulting in some overhead. In out-band D2D communication, different spectra from a cellular network (i.e., industrial, scientific, and medical (ISM) bands) are used; therefore, there is no interference between D2D and cellular communications. Because of this characteristic, D2D and cellular networks communicate simultaneously without any interruption. However, D2D devices may suffer from other network entities that access the ISM band, and QoS is lower compared to in-band D2D owing to the nature of unlicensed bands, limited transmission range, and low data rate.

3GPP started standard activities in Release 12 to enable D2D communication in LTE networks [155]. The D2D standard consists of two parts: device discovery and data communication. The purpose of device discovery is to find other neighboring devices within the transmission range for communication. There are two types of device discovery: type 1 and type 2 (2A and 2B). Type 1 is a collision-based procedure with non-UE-specific allocation. Thus, devices randomly select their radio resources for device discovery in every discovery period. In type 2, the network schedules discovery signal transmission for all UE. In particular, type 2B uses semipersistent allocation only for radio resource control- (RRC-) connected UE with predefined frequency hopping. Similarly, there are two data communication modes: mode 1 and mode 2. In the case of mode 1, an eNodeB allocates data resources to all UE. In particular, in the in-coverage scenario, eNodeB follows the same resource allocation procedure as the cellular mode. In the case of mode 2, a UE assigns its resources from the preconfigured resource pool autonomously. Thus, UE can select and communicate even in out-coverage or partial-coverage scenarios.

*5.3. V2X Communication.* The advent of autonomous cars, high-traffic information systems, and highly reliable safety services has led to the need for a new communication technology for vehicles with high reliability, high data rate, and low latency. This technology is called vehicle-to-everything (V2X) communication, and it includes vehicle-to-vehicle (V2V), vehicle-to-pedestrian (V2P), and vehicle-to-infrastructure (V2I) communication [156]. D2D communication for cellular networks is currently the most suitable option for enabling V2X communication because D2D provides short end-to-end latency and a long transmission range. When V2X communication is deployed in cellular networks, network deployment cost can be reduced and deployment time can be shortened by reusing the infrastructure of a legacy cellular network [18].

To enable road safety services and autonomous driving, vehicles need to exchange several pieces of information. ETSI defines message types for these use cases: cooperative awareness message (CAM) [157] and decentralized environmental notification message (DENM) [158]. CAM is defined for periodical broadcast of short messages to nearby vehicles. This message type delivers presence, position (i.e., GPS information), identifier, and basic status. DENM is transmitted when a specific event such as an accident or abnormal situation occurs to warn neighbor nodes of the event.

V2X communication is currently under standardization by 3GPP since 2015. SA1 defines use cases and service requirements, and SA2 studies network architecture to support use cases in-vehicle networks. RAN is working on radio resource management to satisfy the requirements of vehicular networks. In the US, system requirements for an on-board V2V safety service have been standardized in SAE J2945/1 [159]. ETSI documented Release 1 for a cooperative intelligent safety service message set and Release 2 for urban ITS applications.

V2X communication must support road safety services with high mobility. Therefore, it has very strict requirements compared to other communication technologies to enable highly reliable services. In vehicular networks, all vehicle nodes are moving at high speeds. Thus, local information becomes meaningless rather quickly. Furthermore, V2X communication guarantees the transmission of safety messages and maintains connectivity with neighbors to satisfy reliability requirements. Vehicles should react as fast as possible to accidents. To this end, ultra-low end-to-end delay is required, and a faster duty cycle is needed for communication and device discovery. Figure 13 shows an overview of the bandwidth and latency requirements for V2X communication.

The key performance indicators are as follows [18]:

   (i) E2E delay: 10~100 ms

  (ii) Reliability: $10^{-5}$

 (iii) Positioning accuracy: 30 cm

 (iv) Data rate: 10~40 Mbit/s

*5.4. Challenging Issues and Future Directions in 5G Mobile Access Networks*

*5.4.1. M2M Communication.* The LTE standard was originally targeted to human-to-human (H2H) communication. The M2M communication in cellular IoT should support small data transmission with an irregular time interval. Current radio resource blocks are too large for M2M communication. Thus, new radio resource management schemes are needed to fulfill these requirements of M2M communication. If the M2M mechanism shares radio resource blocks with H2H communication, we will have to minimize impact and interference from M2M communication to H2H communication. The interference management among M2M devices is also a dormant concern due to the massive number of devices in M2M networks.

In M2M communication, cost efficiency is the most important factor owing to the large number of devices deployed in the network. Efficient power management to maximize battery life and network operation is also very crucial. The network capacity should be adequate to handle the massive number of simultaneous connection requests over a wide coverage area. The M2M communication should have capabilities to manage diverse use cases of IoT services. Also, data aggregation and data offloading concepts can be applied to M2M communication to enhance energy efficiency and communication.

*5.4.2. D2D Communication.* When applying D2D communication to legacy cellular networks, we need to manage the interference caused by D2D communication to minimize performance degradation of the cellular network. The network carefully handles the random mobility of UEs and random channel status for interference management. A power control mechanism is mainly used to coordinate the interference in D2D communication. Power control can be achieved either in a centralized way by eNodeB or in a distributed way by each UE. Fundamentally, a centralized approach is more efficient because all the network information including SNR and signaling power of UE is known, but information about additional control and computational overhead is needed to gather network information and adjust the signaling power of all UEs appropriately.

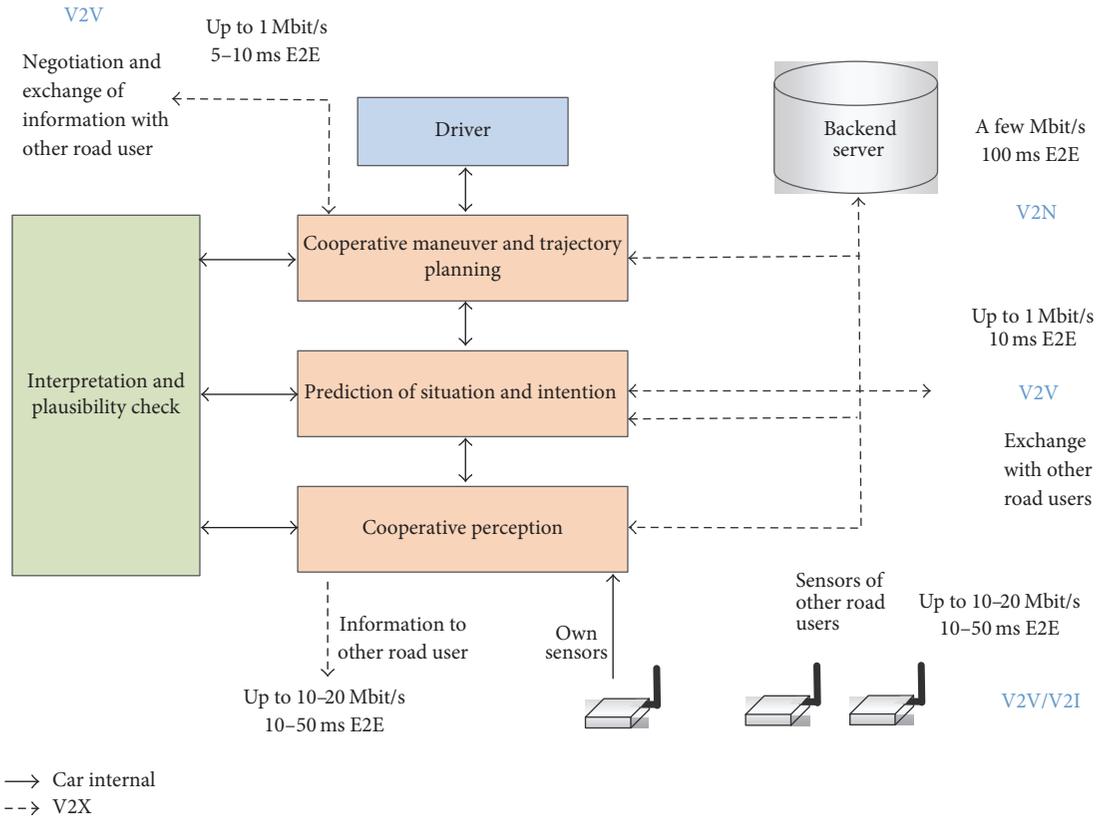

FIGURE 13: Connectivity demands of future connected vehicles [18].

The most challenging issue in D2D communication is interference management. Without efficient interference management, D2D communication cannot coexist with a cellular network. The other issue is mode selection. For example, when the network allows D2D mode, the questions are which event triggers the mode selection procedure, what kind of parameter is used, and whether to use D2D. Data offloading needs to be studied further to increase network capacity and coverage expansion by a relay and power management.

*5.4.3. V2X Communication.* As mentioned earlier, cellular D2D communication can be an enabling technology for V2X communication. However, legacy D2D communication is unable to fulfill the requirements of V2X communication because of its limitations. D2D communication has a collision risk during side-link transmission. Moreover, there are many unsuitable mechanisms leading to latency that is too long for V2X communication, such as long duty cycles for device discovery, inefficient resource allocation and link adaptation, and slow connection setup procedure.

Thus, further studies are needed to apply D2D to V2X. The most important issue is reducing end-to-end latency. Network-assisted radio and link management can be a solution to improve the delay performance. Legacy device discovery mechanisms are too slow for vehicle networks. Thus, new faster device discovery mechanisms should be developed. In addition, a new protocol with a lower duty cycle is needed to minimize the latency and the network control mechanism used in D2D communication should be optimized to reduce control signaling overhead and interference. Additional study issues are supporting flexible retransmission and advanced collision resolution to guarantee reliable requirement and constructing a robust network by increasing link connectivity.

## 6. Conclusion

The expectation of future mobile system or next generation wireless networks comprises high-speed access providing without limitation of time and location. As a consequence, the NGN has to deal with the high data rate, real-time data handling, centralized views of the entire network with minimum delay, greater security, fewer data losses, and less error rate. The development of any technologies with high data traffic and high QoS of universal network infrastructures depends on the integration of new technologies or new services with the existing network infrastructure. In this survey, we have discussed the network architecture, service framework, and topologies that will play an important role to meet the requirements of future networking infrastructure that is 5G network. The requirement of 5G will be massive IoT connectivity, virtual experience and media, and real-time communication. So, the architecture of 5G will be such that the flexibility and scalability of the future network will be maximized. Therefore, the future network will depend on the combination of new technologies such as cloud computing, SDN, NFV, and E2E networking infrastructure. Besides, the integration of SDN with NFV will ensure dynamic data control, centralized network provisioning, and adaptation of new services and innovation. To the best of our knowledge, the survey of promising technologies for 5G networks has emphasized an absolute idea on the interesting attempts in network development trend based on standardization status. Along with this, the promising architectures and services of the SDN, cloud computing, and IoT have been provided. However, the contents presented in this survey are the first step toward the potential architectures and implementation works for 5G network and are not the end picture. After all, the realization of future network for 5G needs a lot of efforts in the research laboratories, industries, and companies.

## Competing Interests

The authors declare that they have no competing interests.

## Acknowledgments

This research was supported by the MSIP (Ministry of Science, ICT and Future Planning), Korea, under the ITRC (Information Technology Research Center) support program (IITP-2016-R2718-16-0004) supervised by the IITP (National IT Industry Promotion Agency).

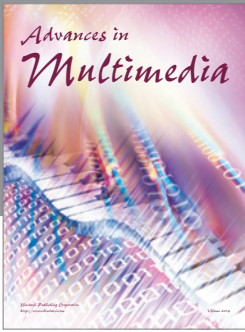
*Advances in*
**Multimedia**

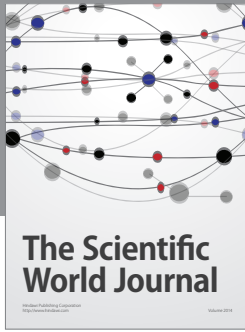
**The Scientific World Journal**

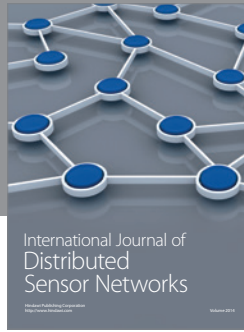
International Journal of
**Distributed Sensor Networks**

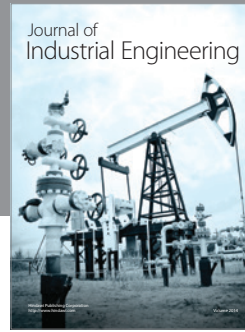
Journal of
**Industrial Engineering**

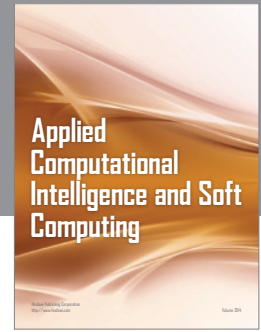
**Applied Computational Intelligence and Soft Computing**

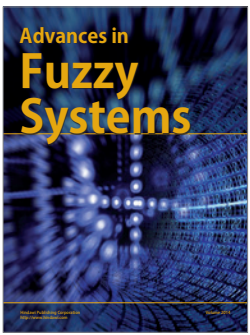
Advances in
**Fuzzy Systems**

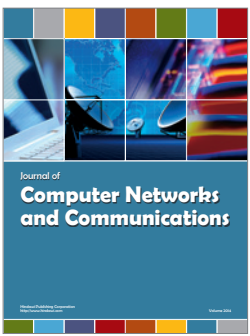
Journal of
**Computer Networks and Communications**

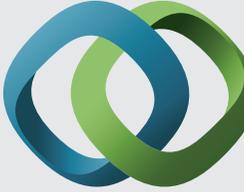
Submit your manuscripts at
http://www.hindawi.com

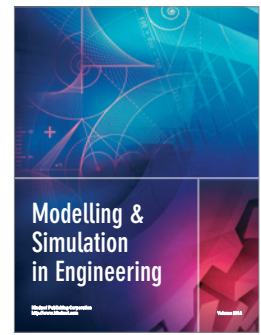
**Modelling & Simulation in Engineering**

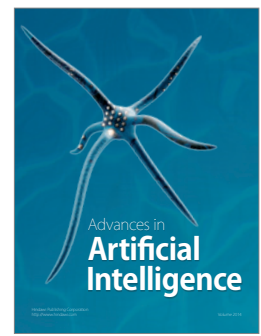
Advances in
**Artificial Intelligence**

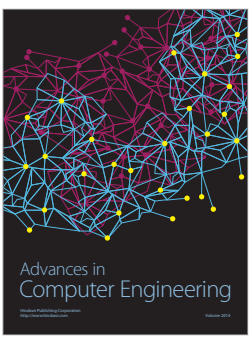
Advances in
**Computer Engineering**

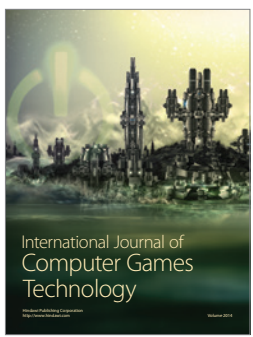
International Journal of
**Computer Games Technology**

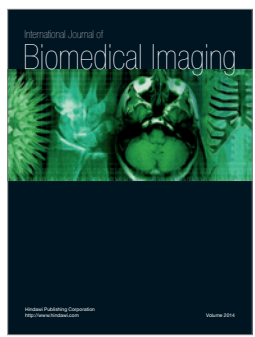
International Journal of
**Biomedical Imaging**

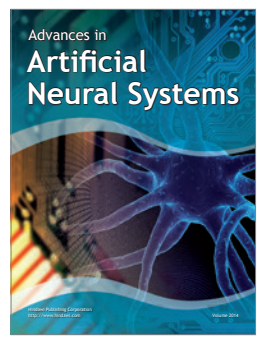
Advances in
**Artificial Neural Systems**

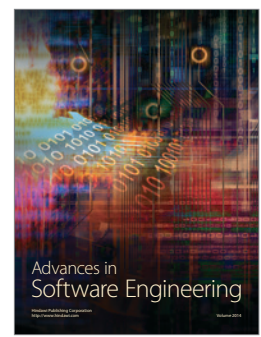
Advances in
**Software Engineering**

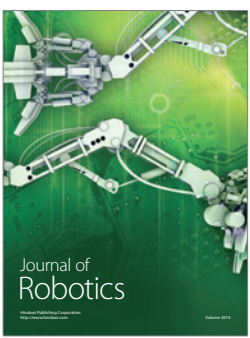
Journal of
**Robotics**

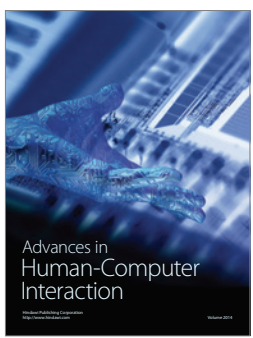
Advances in
**Human-Computer Interaction**

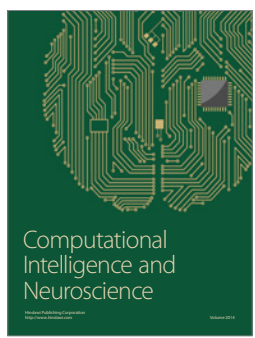
**Computational Intelligence and Neuroscience**

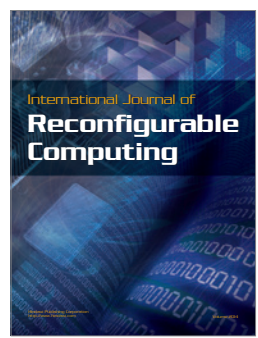
International Journal of
**Reconfigurable Computing**

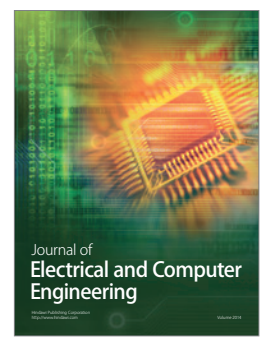
Journal of
**Electrical and Computer Engineering**